\documentclass[10pt,preprint]{aastex}

\usepackage{graphicx,natbib}
\usepackage{amsmath,amssymb}
\newcommand{\esp}{\mathsf{E}}
\newcommand{\var}{\mathsf{var}}

\shorttitle{The Strehl Ratio in Adaptive Optics Images}
\shortauthors{Soummer \& Ferrari}

\begin{document}

\title{The Strehl Ratio in Adaptive Optics Images: Statistics and Estimation}
\slugcomment{ApJL, in press}
\author{R\'emi Soummer \altaffilmark{1}}
\affil{Department of Astrophysics, American Museum of Natural History, 79th Street at Central Park West, New York, NY 10024, USA}

\and

\author{Andr\'e Ferrari \altaffilmark{2}}
\affil{Laboratoire Universitaire d'Astrophysique de Nice, Universit\'e de Nice Sophia Antipolis, Parc Valrose, 06108 Nice, France}
\email{rsoummer@amnh.org, ferrari@unice.fr}


\begin{abstract}
Statistical properties of the intensity in adaptive optics images are usually modeled with a Rician distribution. We study the central point of the image, where this model is inappropriate for high to very high correction levels. The central point is an important problem because it gives the Strehl ratio distribution. We show that the central point distribution can be modeled using a non-central $\Gamma$ distribution.
\end{abstract}


\maketitle 
\section{Introduction}
\label{sect:intro}

In this Letter, we study the statistics of the light intensity at the central point of adaptive optics (AO) images in the case of a high to very high AO correction. This problem corresponds to the statistical properties of the ``instantaneous'' Strehl Ratio (SR) \citep{FC04}. The regime that we consider is relevant to current or future AO systems.

Outside the central point, the statistical properties of the light intensity constrain the detection limits of faint companions to nearby stars. It can be described by a Rician distribution, both in the case of a partial AO correction \citep{CC99c}, and in the case of a high or very high \citep{AS04}. This model has been successfully verified in Lick AO data by \citet{FG06}, and the case of coronagraphic images is detailed in \citet{SFA07}. 

The intensity distribution at the center of AO-corrected images was previously studied by \citet{CC98,CC01} at very low correction levels, for SR in the $10\%-20\%$ regime. They showed a good agreement between the Rician model, numerical simulations \citep{CC99}, and experimental data \citep{CC01b}.
At high to very high AO correction levels, however, the Rician distribution fails to model successfully the statistics of the intensity at the central point. \citet{GCR06} provided evidence of this problem in short exposure Lick AO images, and we further verify this issue using numerical simulations (Sec.\ref{SRstat}).
\citet{GCR06} modeled the observed SR distribution using the statistical properties of the atmospheric seeing. In this paper, we describe an alternative approach based solely on image formation and considering stationary phase properties. The model consists of a non-central $\Gamma$ distribution for the opposite of the modulus of the complex amplitude. Recently, \citet{G06} and \citet{CGR06} independently developed similar approaches to modeling stationary processes, based on a model of the variance of the phase. 

\section{Statistical properties of the intensity}\label{Sec:Rappels}

We briefly recall the derivation of the statistical properties of AO images. A detailed presentation is given by \citet{AS04} and \citet{SFA07}, based on results known in the context of holography by \citet{Goo75,Goo06}.
In this section we use a one-dimensional formalism for clarity, but the results are valid in the general case.
The complex amplitude in the focal plane is the Fourier transform of the pupil plane complex amplitude:
\begin{equation}\label{Eq1}
\Psi(r)=\int P(x) (A+a(x))\, e^{-2\imath \pi x \, r} dx,
\end{equation}
where $P(x)$ denotes the pupil function, normalized such as $\int P(x) dx=1$. $A$ corresponds to the deterministic perfect part of the wave, and $a(x)$ is the zero-mean complex error term: $A+a(x)=e^{\imath \phi(x)}$, with $A=\esp[e^{\imath \phi(x)}]$, where $\esp$ denotes the expectation value. $\phi(x)$ denotes the phase of the wavefront in the pupil plane, assuming a zero-mean Gaussian, with a constant variance $\sigma_\phi^2$. We use the classical definition of SR as the ratio between the central point intensities in the actual and ideal case \citep{Hardy98}. Using the extended Mar\'echal approximation ($SR\approx e^{-\sigma_\phi^2}$), $A$ becomes $A=e^{-\sigma_\phi^2/2}$, which is approximately the square root of SR. 

\emph{Outside the central point of the image}, the distribution of the complex amplitude can be approximated using known results from signal processing. 
Assuming that the complex amplitude $a(x)$ can be represented by discrete values, and that the correlation between two points in the pupil plane decreases with their relative distance, the probability density function (PDF) of the complex amplitude in the focal plane for the random part of Eq.\ref{Eq1} is an asymptotically circular Gaussian \citep{Bri81}. We recall that for a scalar circular Gaussian distribution, i.e. $z \sim \mathcal{N}_c(0,\sigma^2)$, the real and imaginary parts are independent and have the same variance. 
We will consider that for $r\not = 0$, $\Psi$ follows a decentered Gaussian distribution $\Psi \sim \mathcal{N}_c(C,I_s)$, where $C$ is the complex amplitude in the focal plane corresponding to the perfect part of the wave $A$ (see Fig.2 of \citet{AS04}). 
The corresponding intensity follows a Rician distribution:
\begin{equation}\label{Eq:Rician}
p_I(i)=\frac{1}{I_s}\exp\left(-\frac{i+I_c}{I_s}\right)\mathrm{I_0}\left(\frac{2 \sqrt{i} \sqrt{I_c}}{I_s}\right),\; i>0,
\end{equation}
where $I_c=|C|^2$, and $\mathrm{I_0}$ denotes the zero-order modified Bessel function of the first kind. This model has been verified outside the central point of the image in simulations \citep{SFA07} and in real adaptive optics data \citep{FG06}.

\begin{figure}[htbp]
\center
\resizebox{.4\hsize}{!}{\includegraphics{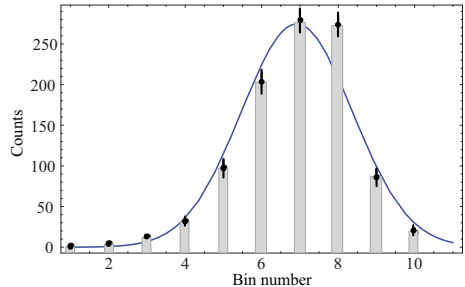}}
 \caption{Histogram of the central point intensity ( SR) for a numerical simulation of 1000 independent PAOLA phase screens with a  SR of $85\%$. The error bars assume statistical Poisson noise. The best fit of a Rician PDF is obtained for $I_c=0.83$, $I_s=3\,10^{-5}$ and is superimposed to the Histrogram. The Rician model fails to describe these data according to the $\chi^2$ and Kolomogorov-Smirnoff tests.}
 \label{Fig:RicianCentralPoint}
\end{figure}

The central point of the image is a particular case, which gives a natural estimator of the  SR. We tested the Rician distribution at this location with a numerical simulation, using PAOLA \citep{JVC06} to generate independent instantaneous intensity values.
We chose the parameters of the AO systems to generate several sets of data with  SR ranging between $50\%$ and $95\%$.
We used a maximum likelihood (ML) estimation of the parameters $I_c$ and $I_s$, assuming Eq.\ref{Eq:Rician}. The Likelihood is computed for the unbinned data and maximized using optimization routines of Mathematica. The starting parameters are obtained from the moments method.
We performed both the $\chi^2$ test and the Kolomogorov-Smirnov (KS) test on these results, using 10 identical bins for the $\chi^2$ test, and a Monte-Carlo estimation of the KS distribution, since the parameters are estimated from the data \citep{Jenkins03}. Both tests conclude that the Rician model is incompatible with the data at the central point, with respective right tail values of $3.8\,10^{-4}$ and $1.6\,10^{-2}$.
This is confirmed by reproducing these tests for a few independent sets of simulated data. We illustrate the histogram of the simulation and the best fit obtained with the Rician model in Fig.\ref{Fig:RicianCentralPoint}.
This result is not surprising since the skewness of the Rician distribution is $(2+3\lambda)(1+\lambda)^{-3/2}$, which is positive, whereas the data histogram clearly exhibits a negative skewness.  
Observations based on real AO data, carried out at Lick Observatory \citep{GCR06} and Palomar Observatory (R. Soummer \& J.P. Lloyd 2007, in preparation), also reported negatively skewed  SR distributions. 

\section{Central point: Strehl Ratio distribution}\label{SRstat}

\begin{figure*}[htbp]
\center
\resizebox{0.8\hsize}{!}{\includegraphics{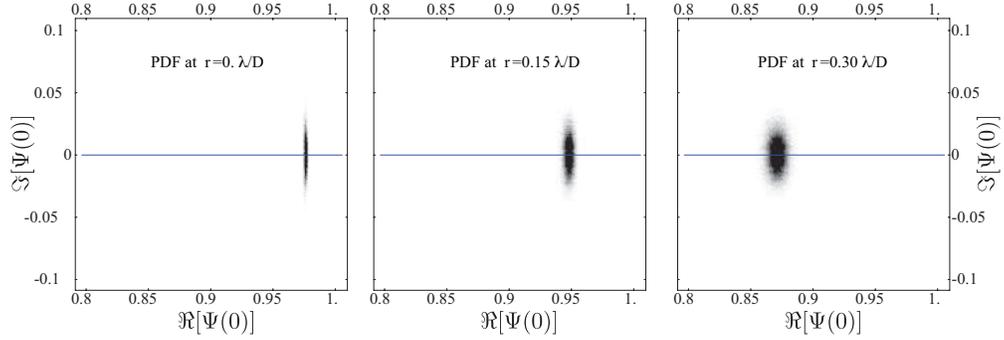}}
 \caption{Histogram of independent realizations of the complex amplitude in the focal plane for a  SR of approximately $95\%$ at three locations indicated in each panel. Left: The complex amplitude at the central point is not a circular Gaussian distribution. When moving away from the center, the distribution becomes progressively circular, and is fully circularized for positions $r>\lambda/D$ (not represented here).}
 \label{FigPDFcenterSR90}
\end{figure*}

\begin{figure}[htbp]
\center
\resizebox{.4\hsize}{!}{\includegraphics{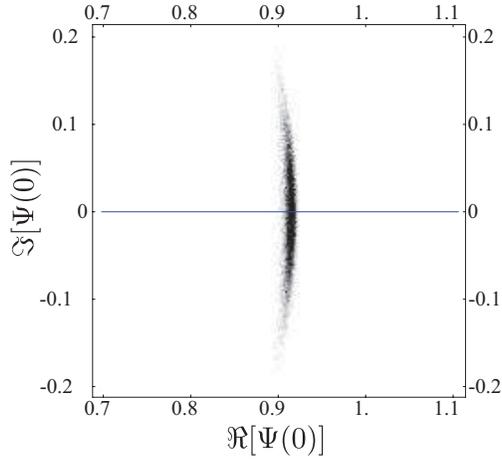}}
 \caption{Histogram of independent realizations of the complex amplitude at the central point in the focal plane, for a  SR of $80\%$. The corresponding distribution of intensity is negatively skewed. This can be seen by looking closely at the distribution of the modulus, along the real axis for example.}
 \label{FigPDFcenterSR80}
\end{figure}

At the center of the image, the Fourier phase term of Eq.\ref{Eq1} vanishes and the complex amplitude is simply the integral of the pupil complex amplitude, $\Psi(0)	=\int P(x)\,e^{\imath \phi(x)} dx$.
At low correction levels, the phase $\phi(x)$ is large, and the vectors $e^{\imath \phi(x)}$ take any orientation in the complex plane. The sum of a large number of these vectors over the aperture produces a random walk, and the corresponding complex amplitude distribution is asymptotically Gaussian, according to a central limit theorem. 
At high correction levels, the phase $\phi(x)$ is small and the vectors $e^{\imath \phi(x)}$ are not oriented randomly in the complex plane. Their sum does not produce a random walk, and the corresponding distribution is not circular Gaussian. This is the case that we study in this section.
When moving outside of the center, however, the Fourier phasors rotate the vectors $e^{\imath \phi(x)}$ in the complex plane, and the sum of the resulting vectors produces a random walk. This explains qualitatively why we obtain a circular Gaussian distribution outside the central point, for $r>\lambda/D$, even at very high correction levels. Fig.\ref{FigPDFcenterSR90} shows the histogram of independent realizations of the complex amplitude, at the central point where it is clearly not circular Gaussian,  and in the transition region $(r\ll \lambda/D)$ where the circularization occurs.
This phenomenon can also be understood qualitatively by considering a Taylor expansion of the phase term $e^{\imath \phi(x)} $ \citep{PSM03}. At a very high  SR, the expansion is mainly dominated by the first oder term $\imath \phi(x)$ and the resulting distribution is an approximately imaginary Gaussian (see Fig.\ref{FigPDFcenterSR90}, left most panel).
Fig.\ref{FigPDFcenterSR80} illustrates the complex amplitude histogram for a  SR of about $80\%$ where the effect of higher order terms is clearly visible, when compared to the left panel of Fig.\ref{FigPDFcenterSR90}. 
At a high or very high SR, we can limit the expansion of $e^{\imath \phi(x)}$ to the second term \citep{SLH02} and we obtain the intensity:
\begin{equation}
I = \left(1 - \int P(x) \frac{\phi(x)^{2}}{2}dx\right)^2 + \left( \int P(x)\phi(x)dx \right)^2
 \end{equation}
The second term corresponds to the mean of the phase residuals over the pupil. It seems reasonable to neglect this term compared to the first term, assuming the phase to be zero mean\footnote{Note that whereas this term can be neglected in the intensity, it cannot be neglected in the second order approximation of the complex amplitude, because it corresponds to the imaginary part, which cannot be neglected compared to the real part.}.
Given that $1 > \int P(x) \frac{\phi(x)^{2}}{2}dx$, we can revert to the complex amplitude:
 \begin{equation}\label{Eq:ApproxMod}
 \Psi(0) \approx 
\left(1 - \int P(x) \frac{\phi(x)^{2}}{2}dx\right) e^{\jmath \varphi},
 \end{equation}
by introducing a random phase term $\varphi$. Note that $\varphi$ is different from the pupil phase $\phi$.

\begin{figure}[htbp]
\center
\resizebox{0.8\hsize}{!}{\includegraphics{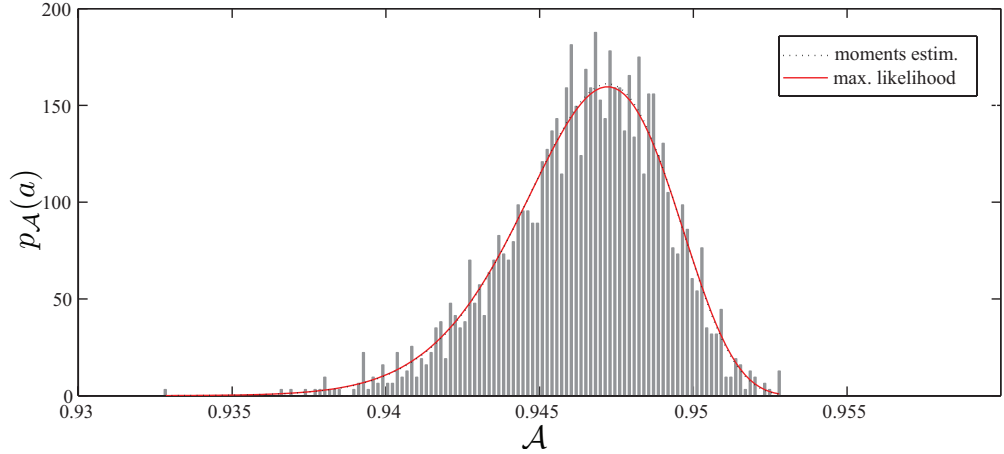}}
 \caption{Numerical simulation for SR$\approx 90\%$. The simulation include both amplitude and phase screens. The figure shows the histogram of $\sqrt{I}$ and the proposed model. Both methods (moments, and maximum likelihood) give almost identical results.}
 \label{figPDFmodelSR90}
\end{figure}

The square of the phase $\phi(x)^2$ is proportional to a $ \chi_1^2$ distributed random variable:
$\phi(x)^{2}= \sigma^2 Q(x)$ where $Q(x) \sim \chi_1^2$. 
The integral $U=\int P(x) \phi(x)^{2}dx $ can be reasonably approximated by the discrete sum: $U=\sigma'^2 \sum_{i=1}^{k}Q(x_i)$, where the number of discrete samples can be assumed as large as necessary, and where $\sigma'^2=\delta\sigma^2$ with $\delta$ the uniform integration step.
The variable $\sum_{i=1}^{k}Q(x_i) \sim \chi_k^2$, where $k$ denotes the degree of freedom of the $\chi^2$ distribution (here corresponding to the number of discrete samples). 
The distribution of $U$ is a $\Gamma$-distribution $\Gamma({k}/{2},2\sigma'^2)$ \citep{JKB94}:
\begin{equation}
p_U(u) = \frac{(2\sigma'^2)^{-k/2} u^{k/2-1} e^{ -u/2\sigma'^2}}{\Gamma(k/2)}, \; u>0.
\end{equation}
The opposite of the modulus of the complex amplitude $\mathcal{A}=U/2-1$ defined as $\Psi(0) = - \mathcal{A}\, e^{\jmath \varphi}$ thus follows a three-parameter non-central Gamma distribution $\Gamma(k/2,\sigma'^2,-1)$.
The approximation of Eq.\ref{Eq:ApproxMod} sets an arbitrary constraint on the mean of the distribution: $\esp(|\Psi(0)|)=1-\sigma_\phi^2/2$, which corresponds to the regime of Mar\'echal's approximation \citep{Hardy98}. The constraint on the mean can be released without changing the statistical model by relaxing the third parameter. Finally, we model $\mathcal{A}$ by the $\Gamma(\alpha,\beta,\gamma)$ distribution defined as:
\begin{equation}\label{gamma}
p_\mathcal{A}(a) = \frac{(a-\gamma)^{\alpha-1}\exp\left( -(a-\gamma)/\beta\right)}{\beta^\alpha \Gamma(\alpha)}, \; \alpha>0,\, \beta>0, \, a>\gamma.
 \end{equation}
The mean and variance of this distribution are:
\begin{equation}\label{Eq:MeanVar}
\esp[\mathcal{A}] = \alpha \beta + \gamma, \;
\var[\mathcal{A}] = \alpha \beta^2.
\end{equation}
The skewness of the distribution is $2/\sqrt{\alpha}$ which is indeed negative for $\sqrt{I}=-\mathcal{A}$.
Estimations of the three parameters can be obtained by a moments method \citep{JKB94}:
\begin{equation}\label{Eq:Moments}
\hat{\alpha} = 4 \frac{\mu_2^3}{\mu_3^2},\;
\hat{\beta} = \frac{\mu_3}{2\mu_2},\;
\hat{\gamma}=\mu_1- \frac{2\mu_2^2}{\mu_3},
\end{equation}
where $\mu_k$ is the classical empirical estimator of the $k$th centered moment.
\citet{JKB94} gives an algorithm for the computation of the ML estimation of $\alpha$, $\beta$, and $\gamma$ together with their asymptotic variance.
The  SR can be expressed directly using the mean and variance of the distribution (Eq.\ref{Eq:MeanVar}):
\begin{equation}\label{Eq:SR}
 SR=\esp[\mathcal{A}^2]=\alpha \beta^2+(\alpha \beta +\gamma)^2.
 \end{equation}
We obtain a new estimator of the  SR by substituting the parameters in Eq.\ref{Eq:SR} with their estimated values, using the moments method or the ML. Using the moments method, the estimation of the  SR is: $SR=\mu_1^2+2\mu_2^2$.
Fig.\ref{figPDFmodelSR90} shows the results of numerical simulations for $SR \approx 90\%$, including 2000 independent realizations of phase and amplitude screens, corresponding to the atmospheric corrected wavefronts and to the atmospheric scintillation. We perform a ML estimation of the three parameters $\alpha$, $\beta$, and $\gamma$, using the moments method to set the initial values. The estimated  SR $(89,58\%)$ from Eq.\ref{Eq:SR}  matches the simulation parameters very well ($SR=89.50\%$).

\begin{figure}[htbp]
\center
\resizebox{0.4\hsize}{!}{\includegraphics{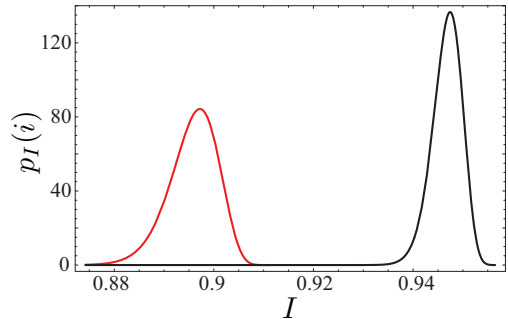}}
 \caption{Examples of two analytical PDF models for the intensity for estimated  SR of $89,58\%$ and $94.50\%$. The skewness of the distribution decreases with  SR and is almost zero at very high  SR.}
 \label{figmodels}
\end{figure}

We can derive an analytical expression for the PDF of the intensity $I=\mathcal{A}^2$, using $p_\mathcal{A}(a)$ from Eq.\ref{gamma}:
\begin{equation}
p_I(i)=\frac{1}{2\sqrt{i}}\left(p_\mathcal{A}(\sqrt{i})+p_\mathcal{A}(-\sqrt{i})\right),\; i>0
\end{equation}
Fig.\ref{figmodels} shows two typical PDF for the intensity, where the parameters have been estimated from simulated data, using the Gamma distribution for the modulus.

\section{Conclusion}
The Rician distribution, which describes successfully the statistics of AO images outside of the central point, is not appropriate for the central point of the image \emph{at high correction levels}. This specific location is particularly important as it corresponds to the SR. 
At high SR levels, the complex amplitude at the central point is not described by a circular Gaussian distribution, which can be explained by the role of Fourier phasors that vanish at the central point and circularize the distribution outside the center. Numerical simulations support this conclusion.

We propose an alternative approach, based on a model of the modulus of the complex amplitude, well described by a non-central Gamma distribution. Our method enables the use of standard algorithms for the estimation of the parameters of this well known distribution. When the parameters have been estimated from the PDF of the modulus, it is possible to revert to the PDF for the intensity by the appropriate transformation. 
Application to real data sets is under study. It would be interesting to compare the approach based on the seeing statistics \citep{GCR06} or on the model of the phase variance \citep{CGR06}, with our results which do not require fluctuations of the seeing.

\vspace{10 pt}

\textit{RS is supported by a Michelson Postdoctoral Fellowship, under contract to the Jet Propulsion Laboratory funded by NASA and managed by the California Institute of Technology, and by an AMNH Kalbfleisch Fellowship. 
This work has been partially supported by the National Science Foundation Science and Technology Center for Adaptive Optics, managed by the University of California at Santa Cruz under cooperative agreement AST 98-76783. The authors thank Gladysz, Christou and Aime for discussions, and Jolissaint for PAOLA.
 }



 \end{document}